\documentclass[12pt]{article}
\usepackage{graphics, color}
\usepackage{graphicx}
\usepackage{amssymb, amsmath, tensor}
\pdfoutput=1

\def\gsim{\, \rlap{$>$}{\lower 1.1ex\hbox{$\sim$}}\,}
\def\lsim{\, \rlap{$<$}{\lower 1.1ex\hbox{$\sim$}}\,}

\begin{document}

%Title page

\begin{titlepage}
\bigskip
\bigskip\bigskip\bigskip
\centerline{\Large  Magnetic $AdS \times {\mathbb R}^2$: Supersymmetry and stability}
\bigskip\bigskip\bigskip
 \centerline{{\bf Ahmed Almuhairi}\footnote{\tt almuhairi@umail.ucsb.edu}}
\medskip
\centerline{\em Department of Physics}
\centerline{\em University of California}
\centerline{\em Santa Barbara, CA 93106}\bigskip
\bigskip
 \centerline{{\bf Joseph Polchinski}\footnote{\tt joep@kitp.ucsb.edu}}
\medskip
\centerline{\em Kavli Institute for Theoretical Physics}
\centerline{\em University of California}
\centerline{\em Santa Barbara, CA 93106-4030}\bigskip
\bigskip
\bigskip\bigskip

%ABSTRACT

\begin{abstract}
We study AdS/CFT with a Kaluza-Klein magnetic field in one plane.  By appropriate choice of magnetic $U(1)$, and by balancing the magnetic field against the background D field, we obtain a supersymmetric field theory.  We find the dual geometry for an $AdS_5 \to AdS_3 \times {\mathbb R}^2$ example, and we compare the moduli spaces and entropies.  For the entropy, the interactions are important even at weak coupling.  We also consider nonsupersymmetric embeddings of the $U(1)$, and show that over a regime of parameter space all known instabilities appear to be absent, aside from a dilaton tadpole that may be removed in a number of ways.

\end{abstract}
\end{titlepage}
\baselineskip = 16pt

\tableofcontents

\section{Introduction}
\setcounter{footnote}{0}

Attempts to model condensed matter systems holographically have been largely phenomenological, assuming a bulk supergravity description without a definite field theory dual.  Construction of top-down models is difficult, both in obtaining specific field theories of interest, and in stabilizing them.  The systems being modeled are non-supersymmetric, which in combination with very strong coupling leads to various instabilities. 
Generally one must begin with a supersymmetric system in the UV, but this includes fundamental scalars; often these  are  not relevant to the physics being modeled but  they have the annoying habit of condensing into the wrong phase.  One needs first to stabilize the moduli.  Even then, there may be BF-forbidden tachyons~\cite{Breitenlohner:1982bm} coming from Kaluza-Klein (KK) modes~\cite{DeWolfe:2001nz}, corresponding to condensation of some composite operator built from the scalars.  There can  also be an instability in which the Coulomb branch for the scalar fields, which is a flat direction in the supersymmetric case, becomes an unstable one.  Of course, one may learn a lot from a state that is long-lived, for example at large $N$, but a fully stable system is more satisfying.

There is a simple class of models that seems to be free of many of these difficulties, where one begins with some relativistic AdS/CFT dual and couples a background magnetic field to the global KK symmetries, e.g.\ some part of the $SO(6)$ of $AdS_5 \times S^5$~\cite{D'Hoker:2009mm,D'Hoker:2009bc,D'Hoker:2010hr,Almuhairi:2010rb}.\footnote{These solutions (and their dyonic versions) had earlier been used as phenomenological duals, beginning in Refs.~\cite{Hartnoll:2007ai,Hartnoll:2007ih,Hartnoll:2007ip,Albash:2008eh}. }   In the gravity description, turning on a magnetic field in one plane of $AdS_D$ causes those two directions to stop contracting in the IR, leading to an $AdS_{D-2} \times {\mathbb R}^2$ geometry there.  In the weak coupling limit, there is a large degeneracy of excitations in the lowest Landau level (LLL), whose dynamics are effectively reduced by two dimensions.  Thus it is plausible that the effective quantum field theory of the LLL is dual to the $AdS_{D-2}$ region of the geometry.

The stability of these solutions is promising.  Ref.~\cite{Almuhairi:2010rb} showed that for the simplest embedding of the $U(1)$ there are no BF-forbidden KK tachyons (within the supergravity truncation), and actually no tachyons at all.  The magnetic field should also tend to stabilize the Coulomb branch through the inverse of the Meissner effect: condensation of a charged field expels a magnetic field, so introducing a magnetic field should prevent the condensation.  In this paper we will explore these issues further.

In order to get some insight we will first study a supersymmetric example.  The system is again just $AdS_5 \times S^5$, but with a condition on the embedding of the magnetic $U(1)$ in $SO(6)$, and with the auxiliary field of the background gauge multiplet also turned on.  As just discussed, supersymmetry may not be necessary for stability, and further it will bring in features not wanted for modeling real systems, but it is helpful to look first at situations where there is a greater amount of theoretical control.  In particular, the absolute stability is useful in debugging the more general calculation.

In \S2 we show that supersymmetry can be preserved in the field theory by turning on an auxiliary D field along with the magnetic field.  In \S3 we find, for the $AdS_5 \times S^5$ theory, the dual geometry.  In \S4 we study the Coulomb branch of the bulk theory in order to help identify the dual field theory, and as a prelude to the nonsupersymmetric case.  In \S5 we compare the entropies on the two sides, and find that the interactions must play a nontrivial role even in the weak coupling limit.\footnote{After a prior version of this work, Benini and Bobev~\cite{Benini:2012cz,Benini:2013cda} have matched the central charges using $c$-extremization.}
In \S6 we study various instabilities.  For most of these the absolutely stable supersymmetric theories are continuously connected to a parameter space of stable nonsupersymmetric theories, with overlapping regions of stability.\footnote{A prior version of this analysis contained errors which have been found and corrected by A.~Donos, J.~P.~Gauntlett and C.~Pantelidou~\cite{Donos:2011pn}; the qualitative results are not changed.} 
  The one exception is the dilaton, which will generically 
develop a tadpole (except at the self-dual point).  Thus, construction of fully stable solutions will require additional model-building, or a starting point such as M theory that has no dilaton.  We also show that the supersymmetric value of D is an RG attractor.  In \S7 we conclude, including a discussion of the extension to $AdS_2$.

\section{Supersymmetric magnetic field theories}
\setcounter{equation}{0}

The near-horizon limit of any four-dimensional BPS magnetic black hole gives a supersymmetric $AdS_2 \times S^2$ geometry.  However, the spacetime and magnetic curvatures are not independent, so we cannot scale to $AdS_2 \times {\mathbb R}^2$. To see that such geometries are possible we start on the field theory side.  Introducing a $d=4, \,{\cal N} = 1$ or $d=3,\, {\cal N} = 2$ superfield for the  background gauge multiplet, the variation of the gaugino is
\begin{equation}
\delta\lambda = (F_{ij} \sigma^{ij} + i {\rm D}) \epsilon \,;
\end{equation}
we use a Roman font for the auxiliary field in order to distinguish it from the covariant derivatives.
Provided that the background values of the auxiliary and magnetic fields are related by ${\rm D} = \pm 2B$, there will be a zero eigenvalue and two unbroken supersymmetries.  By convention we take $B>0$, but leave the sign of ${\rm D}$ general.  These fields are fixed and do not have to satisfy any equations of motion.  This same approach was recently applied to supergravity backgrounds in Ref.~\cite{Festuccia:2011ws}.

To see the effect of this background, couple it to a single massless chiral multiplet of charge $q$.  The field equation for a massless scalar is 
\begin{equation}
(-\partial_0^2  + \partial_3^2 + D_+D_- + D_- D_+ - q{\rm D}/2) \phi = 0 
\end{equation}
where $D_{\pm} = (D_1 \pm i D_2)/\sqrt2$ and $D_i = \partial_i - i q A_i$.
For a $2+1$ dimensional theory the $\partial_3^2$ term is absent.
Here the gauge-covariant derivatives satisfy $[D_+,D_-] = -qB$.  The lowest Landau level (LLL) solutions satisfy $D_\pm \phi = 0$ where the sign matches that of $q$, and then $D_\mp$ acts as a raising operator.  On the $n$th level,
\begin{equation}
- D_+D_- - D_- D_+ + q{\rm D}/2 = (2n + 1)|q|B + q{\rm D}/2 \,,
\end{equation}
which is an effective $1+1$ dimensional mass-squared in $3+1$ dimensions, and an energy-squared in $2+1$ dimensions.  Applying the same to the Dirac equation,
\begin{equation}
(\gamma^0 \partial_0 + \gamma^1 \partial_1 + \gamma^+ D_+ + \gamma^- D_-)\psi = 0 \,,
\end{equation}
gives the spectrum
\begin{equation}
m^2 = 2n |q |B \,.
\end{equation}
In 3+1 dimensions the $n = 0$ level is left-moving for $q <0$ and right-moving for $q >0$.   The degeneracy of each level is $\nu = |q| N_\Phi$, where $N_\Phi = BV_2/2\pi$ and $V_2$ is the volume of the magnetic plane.

In the absence of the D-field background, the scalar and fermion spectra are out of alignment, but for the supersymmetric value the massive levels are exactly aligned.  For $q{\rm D}>0$ the massless spectrum consists a chiral fermion, moving in a direction depending on $q$ as above, while for $q{\rm D}<0$ it consists of a scalar and a chiral fermion, which for given ${\rm D}$ moves in the opposite direction.  These are the fermion and chiral multiplets of $(0,2)$ supersymmetry.

To apply this to the $d=4,\, {\cal N}=4$ theory, we can gauge any $U(1) \in SO(6)$, but to obtain a supersymmetric result it must be in the $SU(3)$ that leaves some supercharge invariant.  Writing the theory in terms of three chiral multiplets, the superpotential 
\begin{equation}
{\rm Tr}(\Phi_1 [ \Phi_2,\Phi_3])  \label{n4supot}
\end{equation}
 implies that $q_1 + q_2 + q_3 = 0$.  If any charge vanishes there is a doubling of the supersymmetry, but as we will note in the next section there is then also no AdS solution at low energy.  Thus we take all charges nonvanishing, say $q_{1,2} > 0$ and $q_3 < 0$.  Then for ${\rm D}<0$, $\Phi_1$ contributes $\nu_1 = |q_1| N_\Phi N^2$ chiral multiplets, $\Phi_2$ contributes $\nu_2 = |q_2| N_\Phi N^2$ chiral multiplets, and $\Phi_2$ contributes $\nu_3 = \nu_1 + \nu_2$ fermion multiplets, so the total spectrum is nonchiral.  For ${\rm D}>0$ the chiral and fermionic multiplets are reversed, but we will see that in this case there is not an AdS dual.

To be specific we will stay with the convention that $q_{1,2} > 0$ and $q_3 < 0$, but the physics depends only on the magnitudes $|q_1|$, $|q_2|$, and $|q_3|$: if an even number of signs are flipped we can take one of the other four symmetries, and if an odd number are flipped we can take the opposite sign for \mbox{D}.  In particular, the condition for supersymmetry is that these three magnitudes exactly saturate the triangle relation.  For nonsupersymmetric theories, stability will depend in part on whether the magnitudes satisfy the triangle relation.   

\section{Supergravity description}
\setcounter{equation}{0}

Coupling the gauge theory to background gauge and $\rm D$ fields corresponds to a perturbation of the boundary conditions of the dual spacetime~\cite{Maldacena:1997re,Gubser:1998bc,Witten:1998qj}.  The gauge background maps directly to the boundary condition on the KK gauge fields, while the $\rm D$ background is coupled to the operator $-\frac12 {\rm D}\sum_I q_I\Phi_I^* \Phi_I$, which corresponds to an $L=2$ deformation of the $S^5$.  The 10-dimensional solution can then be constructed from the $D=5$ $U(1)^3$ truncation of gauged supergravity~\cite{Cvetic:1999xp}.

More generally, consider supersymmetric solutions to $U(1)^m$ gauged supergravity~\cite{Gunaydin:1984ak},
\begin{eqnarray}
0 = \delta\psi_M &=& \left( {\cal D}_M + \frac{i}{8} X_I (\Gamma_M\!^{NP} - 4 \delta_M\!^N \Gamma^P) \tilde F^I_{NP} + \frac{1}{2} g \Gamma_M X^I V_I \right) \epsilon \,, \nonumber\\
0 = \delta\lambda_i &=& \left( \frac{3}{8} \Gamma^{MN} \tilde F^I_{MN} \partial_i X_I + \frac{3i}{2} g V_I \partial_i X^I - \frac{i}{2} {\cal G}_{ij} \Gamma^M \partial_M \phi^i
\right) \epsilon \,, \label{susyvar}
\end{eqnarray}
where the covariant derivative is 
\begin{equation}
{\cal D}_M \epsilon = \left( \partial_M + \frac{1}{4} \omega_{MNP} \Gamma^{NP} - \frac{3i}{2} g V_I \tilde A^I_M \right) \epsilon\,.
\end{equation}
  The $X^I$ are $m$ scalars subject to one constraint, while the $\phi^i$ parameterize these in terms of $m-1$ nonredundant scalars.  We are using the notation of Ref.~\cite{Cacciatori:2003kv}, with one exception.  In supergravity it is conventional to include a factor of $g$ (the inverse $AdS_5$ radius) in the covariant derivative, but it is not natural to include this supergravity scale in the field theory derivative.  We therefore distinguish these, using a $\tilde{\ }$ for the supergravity gauge fields.  Thus, $g\tilde B = B$.

In order for the plane of the magnetic field to be flat we need the last term in the covariant derivative to vanish,
\begin{equation}
V_I \tilde A^I_M = 0 \,. \label{neutral}
\end{equation}
For $B^I = q_I B$ this is again the condition $\sum_I q_I = 0$.

This allows for various top-down constructions.  For the $d=4,\,{\cal N} = 4$ theory, the $U(1)^3$ $S^5$ truncation of IIB supergravity~\cite{Cvetic:1999xp} has $V_1 = V_2 = V_3 = 1/3$, $X^1 X^2 X^3 = 1$, and $X_I = 1/3X^I$.  In this case there are three other supersymmetries, obtained by flipping the relative signs of the gauge terms with different $I$, consistent with the observation on the field theory side that the physics depends only on the magnitudes of the chrages.
Generically the condition~(\ref{neutral}) can only hold for one of the four supersymmetries at a time.  In the special case that one of the three $\tilde A^I_M$ vanishes the potential supersymmetry is doubled.

We take the Ansatz
\begin{eqnarray}
ds_5^2 &=& e^{2W(y)} \eta_{\mu\nu} dx^\mu dx^\nu + e^{2U(y)} dx^\alpha dx^\alpha + dy^2 \,, \quad \mu,\nu \in 0,1 \,, \ \alpha,\beta \in 2,3 \,, \nonumber\\
\tilde F^I_{23} &=& q_I \tilde B/g \,, \quad \sum_I q_I = 0\, ,
\end{eqnarray}
with the scalars $\phi^i$ also depending only on $y$.
The condition for a rigid supersymmetry is the vanishing of the variations~(\ref{susyvar}) with $\partial_\mu \epsilon = \partial_\alpha\epsilon = 0$:
\begin{eqnarray}
\Gamma_\mu ( 4 W' \Gamma^y + i \Gamma^{\alpha\beta} X_I \tilde F^I_{\alpha\beta} + 4 g V_I X^I ) \epsilon &=& 0
\,, \nonumber\\
( U' \Gamma_\alpha \Gamma^y - i \Gamma^{\beta} X_I \tilde F^I_{\alpha\beta} + g \Gamma_\alpha V_I X^I) \epsilon
 &=& 0\,, \nonumber\\
 ( 8 \partial_y + i \Gamma_y \Gamma^{\alpha\beta} X_I \tilde F^I_{\alpha\beta} + 4 g \Gamma_y V_I X^I) \epsilon &=& 0
\,, \nonumber\\
( 3 \Gamma^{\alpha\beta} \tilde F^I_{\alpha\beta} \partial_i X_I + 12 i g V_I  \partial_i X^I
- 4i {\cal G}_{ij} \Gamma^y \phi^{i\prime}
) \epsilon
 &=& 0\,. \label{rigid}
\end{eqnarray}
To solve these we need the supersymmetry to be an eigenspinor,
\begin{equation}
\Gamma_y \epsilon = - \epsilon \,, \quad \Gamma^1 \Gamma^2 \epsilon = \pm i e^{-2U} \epsilon \,,
\end{equation}
so that two of the eight real supersymmetries can be unbroken. The sign under $\Gamma_y$ corresponds to a choice of direction for the $y$ coordinate.
The conditions~(\ref{rigid}) become
\begin{eqnarray}
 ( 2 W' \pm e^{-2U} \tilde B q_I  X_I  - 2 g V_I X^I ) \epsilon &=& 0
\,,  \label{mucond}\\
( U' \mp e^{-2U} \tilde B  q_I  X_I - g V_I X^I ) \epsilon
 &=& 0\,, \label{alcond}\\
 ( 4 \partial_y \pm e^{-2U} \tilde B q_I   X_I  - 2 g V_I X^I ) \epsilon &=& 0
\,, \label{ycond}\\
(\pm  3 e^{-2U} \tilde B q_I  \partial_i X_I + 6 g V_I \partial_i X^I
+ 2 {\cal G}_{ij} \phi^{i\prime}
) \epsilon
 &=& 0\,. \label{lamcond}
\end{eqnarray}

By combining conditions~(\ref{mucond}, \ref{ycond}) we obtain $\epsilon(y) = e^{W(y)/2}\epsilon_0$.  The remaining conditions give first-order equations for the metric components and scalars.  These describe a flow from the $AdS_5$ UV geometry to the geometry that describes the IR physics.  We are  interested in nontrivial IR fixed points, so we consider  the case that $U' = \phi^{i\prime} = 0$, for which the remaining conditions can be rewritten
\begin{eqnarray}
\pm e^{-2U} \tilde B  q_I  X_I + g V_I X^I &=& 0 \,, \label{e2u}\\
\partial_i ({[V_I X^I ]^2}/{q_J X_J}) = 0 \,, \label{supot}\\
2W' - 3 g V_I X^I = 0 \,.
\end{eqnarray}
The first of these determines the metric component $U$ in terms of the scalars, the second determines the scalars in terms of stationarity of an effective superpotential, and the third gives an $AdS_3$ geometry with radius 
\begin{equation}
L_3^{-1} = 3 g V_I X^I/2 \,.  \label{radius}
\end{equation}

Specializing to $d=4,\,{\cal N} = 4$, Ref.~\cite{Almuhairi:2010rb} found from the field equations that
\begin{eqnarray}
X^I &=& \tilde q_I^2 \,, \quad \tilde q_I \equiv \frac{q_I}{|q_1 q_2 q_3|^{1/3} }\,, \nonumber\\
 e^{2U} &=& \frac{B|q_1 q_2 q_3|^{1/3}}{2g^2} \,, \nonumber\\
 L_3^{-2} &=& g^2 |q_1 q_2 q_3|^{2/3} \sum_I \frac{1}{q_I^2} \,,
\label{xsol}
\end{eqnarray}
and with some manipulation one can show that this solution satisfies the supersymmetry conditions~(\ref{e2u}, \ref{supot}) when $q_1 + q_2 + q_3 = 0$.  We need the lower sign in Eq.~(\ref{e2u}) with our convention that one $q_I$ is negative.  Note that the solution degenerates if any of the $q_I$ vanish, so there is no AdS solution when the supersymmetry is doubled.

\section{Coulomb branch}
\setcounter{equation}{0}

In \S 2 we saw that a given set of charges was associated to two distinct field theories, according to the sign of the D field.  However, we have found only one supergravity solution for each set.  In order to identify the field theory, it is useful to look at the Coulomb branch.  This is a bit subtle, but the result will also be used in studying the stability of the nonsupersymmetric example.

First consider the naive Coulomb branch, a brane whose transverse coordinates are constant.  Combining the uplifted metric~\cite{Cvetic:1999xp} 
\begin{equation}
ds_{10}^2 =  \tilde\Delta^{1/2} ds_5^2 + \ldots \,,\quad \tilde\Delta = \sum_I X^I \mu_I^2
\end{equation}
with the relevant part of the 4-form potential,
\begin{equation}
C_{0123} = g L_3 e^{2y/L_3} e^{2U} \sum_I \left( \mu^2_I (X^I)^2 - X^I \Tilde\Delta \right) + \ldots
\end{equation}
gives for the potential energy density of a probe D3-brane
\begin{equation}
{\cal E} = e^{2y/L_3} e^{2U}  \tilde\Delta^{1/2} + C_{0123} \,. \label{coul1}
\end{equation}
Using the solution~(\ref{radius}, \ref{xsol}) for a probe moving in the $K$ plane ($|\mu_I | = \delta_I {}^K$) gives
\begin{equation}
{\cal E} = e^{2y/L_3} e^{2U} \frac{ X^K}{\sum_I X^I} \biggl(2X^K - \sum_I X^I \biggr)  =
\frac{e^{2y/L_3} e^{2U}}{\sum_I X^I} \frac{ 2 q_K q_1 q_2 q_3}{|q_1 q_2 q_3|^{4/3}} \,.  \label{pot1}
\end{equation}
This is negative for the two $q_K$ of like sign, and positive for the third, confirming the sign asserted for D in \S2.  Evidently the other supersymmetric field theory does not have a strongly coupled conformal fixed point.  We will conjecture as to why this should be in \S5.

There is also a positive energy density from the diamagnetic terms in the covariant derivatives, which diverges with volume and stabilizes the negative case above.  However, our assumption that the transverse coordinates are constant is not gauge invariant, and so we should be more general and allow these to depend on the magnetic plane coordinates $x^\alpha$.

To do this we first rewrite the 10-dimensional metric of Ref.~\cite{Cvetic:1999xp} in a convenient coordinate system.  In terms of the Poincar\'e coordinates $x^\mu, x^\alpha$ and three complex transverse coordinates $z^I= e^{gyX^I} \!\mu_I e^{-i\phi_I} $, it is
\begin{eqnarray}
ds_{10}^2 &=& \tilde\Delta^{1/2}( e^{2y/L_3} \eta_{\mu\nu} dx^\mu dx^\nu + e^{2U} dx^\alpha dx^\alpha )
+  \frac{1}{g^2 \tilde\Delta^{1/2}}  \sum_I \frac{e^{-2gy X^I}}{X^I} |Dz^I|^2\,, \nonumber\\
\tilde\Delta &=& \sum_I X^I | z^I |^2 e^{-2gy X^I} \,,
\quad Dz^I = dz^I + igq_I \tilde A z^I \,, \quad
1 = \sum_I |z^I|^2 e^{-2gyX^I} \,.
\end{eqnarray}
In particular, the transverse space is complex.  The coordinate $y$ is implicitly determined by the last condition.

The potential energy terms from the DBI action are then
\begin{equation}
- \int d^4x \, \tilde\Delta e^{2y/L_3} e^{2U} \det\!^{1/2}\!\left(\delta_{\alpha\beta} + \frac{1}{g^2 \tilde\Delta e^{2U}}   
\sum_I  \frac{e^{-2gy X^I}}{X^I}  D_{(\alpha} z^{I*} D_{\beta)} z^I \right) 
\end{equation}
for general $z^I\!(x^\alpha)$.
The energy should be lowest for an LLL configuration, i.e. $D_{\pm} z^I = 0$ with the sign being that of $q_I$, and again for simplicity we assume that only one of the $z^K$ is nonzero.
The DBI action becomes
\begin{equation}
\int d^4x \, e^{2y/L_3} \left( \tilde\Delta e^{2U} + \frac{1}{2 g^2}  \frac{e^{-2gy X^K}}{X^K}  D_{\mp} z^{K*} D_{\pm} z^K \right) \,.
\end{equation}
The second term can be rewritten
\begin{equation}
\Delta S_{\rm DBI} = -\frac{L_3^2 X^K}{2} \int d^4x \, D_{\pm} (z^K)^{1/gL_3X^K} D_{\mp} (z^{K*})^{1/gL_3X^K}
= \frac{L_3 }{2g}  |q_K| B  \int d^4x \, e^{2y/L_3} \,;  \label{coul2}
\end{equation}
note that $(z^K)^{1/gL_3X^K}$ has charge $1/gL_3 X^K$ in the covariant derivative.  Using the solution (\ref{xsol}), the energy density becomes 
\begin{equation}
{\cal E} = e^{2y/L_3} e^{2U} \frac{ |q_K| X^K}{ \sum_J q_J/X^J} =
\frac{e^{2y/L_3} e^{2U}}{\sum_I X^I} \frac{ 2 |q_K|}{|q_1 q_2 q_3|^{1/3}}\,. \label{pot2}
\end{equation}

The two contributions~(\ref{pot1}, \ref{pot2}) are equal up to a sign, which is opposite if $q_K$ is one of the two positive charges.  Thus there are Coulomb branches in two of the transverse planes, matching the spectrum of bosons found at weak coupling in \S2.  In the third plane the terms add and there is a potential.  The branches are analytic function spaces, corresponding to the LLL wavefunctions.  One can think of the scalar field along them as a configuration of vortices.

\section{Central charge}
\setcounter{equation}{0}

For this $AdS_3$ system the relevant quantity is the central charge rather than the zero point degeneracy.  On the supergravity side this is given by~\cite{Brown:1986nw,Strominger:1997eq}
\begin{equation}
c_{\rm sugra} = \frac{3L_3}{2G_3} =  \frac{3L_3 V_2 e^{2U}}{2G_5} = \frac{3L_3 V_2 N^2 g^3e^{2U}}{\pi} \,,
\end{equation}
using the standard $AdS_5 \times S^5$ result $G_5 = G_{10} g^5 / \pi^3 = \pi/2 N^2 g^3$.  For the solution in \S3 this becomes
\begin{equation}
c_{\rm sugra} = 12 N_\Phi N^2 (q_1 q_2 q_3)^2\frac{\sum_I 1/q_I}{\Bigl(\sum_I q_I^2\Bigr)^2} = 
\frac{ 6 N_\Phi N^2 |q_1 q_2 q_3|}{\sum_I q_I^2} \,. \label{sugrac}
\end{equation}
In \S2 we found in the free limit that
\begin{equation}
c_{\rm free} = 3|q_3|  N_\Phi N^2\,,
\end{equation}
where again $q_3$ is the charge of opposite sign to the other two.
These do not agree.  For example,
\begin{eqnarray}
q_1 = q_2 = 1\,\ q_3 = -2:&&\quad c_{\rm sugra} = 2 N_\Phi N^2\,,\quad c_{\rm free} = 6 N_\Phi N^2 \,, \nonumber\\
q_1 = 1\,,\ q_2 = q \gg1\,,\  q_3 = -1-q:&&\quad c_{\rm sugra} \to 3 N_\Phi N^2\,,\quad c_{\rm free} \to  3q N_\Phi N^2
\,.
\end{eqnarray}

The central charge should be invariant under continuation between weak and strong coupling, both because this is generally true for moduli spaces of CFT's, and because it is related to the $U(1)$ anomaly in the $(0,2)$ algebra.  However, the weakly coupled theory does not approach the free theory, because there is a relevant interaction.\footnote
{More precisely, if we take the coupling to zero while focusing on a given energy scale, then the free theory governs the limit.  If we go to zero energy at any fixed coupling, the interactions become strong.}  
In the low energy theory we have $|q_1| N_\Phi N^2$ chiral multiplets $\Theta_{1i}$, $|q_2| N_\Phi N^2$ chiral multiplets $\Theta_{2j}$, and $|q_3| N_\Phi N^2$ fermionic multiplets $\Psi_{3k}$.  Inserting these modes into the UV superpotential~(\ref{n4supot}) induces a low energy $(0,2)$ superpotential
\begin{equation}
c_{ijk} \Theta_{1i} \Theta_{2j} \Psi_{3k} \,,
\end{equation}
where $c_{ijk}$ is from the overlap of LLL wavefunctions.  Thus, the effective field theory of the supersymmetric LLL is a (0,2) Landau-Ginsburg (LG) model.  

Following the first version of the present work, Benini and Bobev have shown that the central charge can be calculated by a procedure of $c$-extremization~\cite{Benini:2012cz,Benini:2013cda}.  They find that the field theory result for the central charge agrees with the supergravity value~(\ref{sugrac}).

We will discuss one puzzle, which was raised by E. Silverstein.  If we look at the configurations on the Coulomb branch where $\Theta_1$ is nonzero, then $\Theta_2$ and $\Psi$ are massed, and we have $|q_1|  N_\Phi N^2$ chiral superfields of total central charge
$3|q_1|  N_\Phi N^2$; similarly the $\Theta_2$ branch has total central charge $3|q_2|  N_\Phi N^2$.  The greater of these would seem to be a lower bound on the total $c$, but it exceeds the strong coupling value that we have determined.

To resolve this, note that the dimension of these branches would usually be reduced to order $N$ by the $U(N)$ D-term potential.  Here, the $U(N)$ gauge multiplet does not couple to the background field, and so has no LLL states.  Thus we have treated it as decoupled from the low energy theory.  To see how this works for the D-term (which is similar to other decoupling situations), note that the $U(N)$ current is
\begin{equation}
[\Phi_1^*, D_+ \Phi_1] - [D_+  \Phi_1^*, \Phi_1]  = - \partial_+ [\Phi_1^*, \Phi_1]
\end{equation}
(using the LLL condition), and similarly for the $-$ current.  The exchange of $A_\pm$ is then proportional to
\begin{equation}
\frac{k_+ k_-}{-\omega^2 + k_1^2 + 2k_+ k_-}
\end{equation} 
This has just the right form to cancel a local $-1$ from the $U(N)$ D-field, leaving a remainder of order $-\omega^2 + k_1^2$: the potential is reduced to a nonlinear kinetic term, which is related to a four-Fermi interaction by supersymmetry.  This is presumably irrelevant at the LG point, but marginal on the $A_{1,2}$ branches.  We conjecture that it runs to strong coupling and gaps out the off-diagonal terms (due to the commutator structure), leaving a $c$ of order $N$.  This is consistent with the strong coupling analysis of \S4, which described a Coulomb branch of dimension $N$ (times $|q_K| N_\Phi$).  

Consider now the other sign for the background D-field, for which we have not found a strong coupling dual.  In this case the low energy superfields are $\Psi_{1i}$, $\Psi_{2j}$, and $\Theta_{3k}$, and so no low energy superpotential is induced.  The leading interactions are the classically marginal interactions noted above.  If these have the conjectured effect above then they gap the non-Abelian dynamics and leave just the moduli space dynamics, without a geometric dual.

Without supersymmetry, we expect to have only fermions: the IR AdS geometry should be dual to a purely fermionic theory.  As in the above examples, the LLL dynamics will depend on the specific form of the four-Fermi interaction that descends from the gauge theory.  It seems challenging to determine the precise form of this interaction, which should be a function of the underlying gauge theory coupling.

\section{Stability}
\setcounter{equation}{0}

We now consider various possible modes of instability, not assuming supersymmetry.   Note that classically we can adjust both the $U(1)$ charges and the D field, but in the quantum theory the latter runs to a fixed point and only the charges can be varied.

\subsection{Neutral modes and RG flow}

We consider first the $U(1)^3$-singlet fluctuations of the various supergravity scalars.  The potential energy terms in the $U(1)^3$ truncation~\cite{Cvetic:1999xp} are 
\begin{equation}
{\cal U} = - 4 g^2 e^{-4U} \sum_I \frac{1}{X^I} + \frac{ B^2}{2g^2} e^{-8U} \sum_I \frac{q_I^2}{(X^I)^2} \,.
\end{equation}
We have included the metric component $U$ from reduction to 3 dimensions, and transformed to 3-dimensional Einstein frame.  The $X^I$ are constrained by $X^1 X^2 X^3 = 1$, but we can conveniently combine these with $U$ to form unconstrained fields $Y_I = e^{-4U}/X^I$.  The potential is then
\begin{equation}
{\cal U} =   \frac{ B^2}{2g^2}  \sum_I  {q_I^2}\biggl(Y_I - 4 \frac{g^4}{B^2q_I^2} \biggl)^2  - \frac{8g^6}{B^2} \sum_I  \frac{1}{q_I^2} \,. 
\end{equation} 
The solution~(\ref{xsol}), equivalent to $Y_I = 4 {g^4}/{B^2q_I^2}$, is immediate, and we see that this is a global minimum for these fields.

This result also has implications for stability under RG flow.
The fields $X_I$ are dual to the bilinear scalar operator to which the background D couples.  The positive mass-squared for these fields implies that this perturbation is irrelevant in the IR.  That is, we do not need to tune the D field to the exact value required for supersymmetry --- the IR physics will be supersymmetric if we start close enough.  It is only necessary to choose the charges to sum to zero, with any choice of signs.   It would be interesting to solve the full field equations and determine the domain of attraction for this fixed point.  It cannot be the entire range $-\infty <  D < \infty$: if we start with the value that preserves the opposite supersymmetry and has no AdS dual, we cannot flow to this one.

\subsection{Charged modes}

The $SO(6)$ reduction~\cite{Cvetic:2000nc} describes a larger set of 20 scalars in terms of a real symmetric unimodular matrix $T_{ij}$, where $i,j = 1,\ldots,6$.  These have charges $\pm q_I \pm' q_J$ for $I,J \in 1,2,3$.  They include the neutral modes discussing in \S6.1, plus 18 charged modes.
Another possible instability~\cite{Ammon:2011je} comes from bulk gauge fields charged under the magnetic $U(1)$.  These modes have the same gauge charges as the KK scalars, and in fact they mix.  In an earlier version of this work this mixing was overlooked.   

This oversight was noted by Donos, Gauntlett, and Pantelidou~\cite{Donos:2011pn}, who completed the analysis.  They find a narrow region of stable nonsupersymmetric solutions.  Satisfying the triangle inequality is found to be necessary, but not sufficient, for stability; in fact, the simplest case~\cite{D'Hoker:2009mm} $q_1 = q_2 = q_3$ is unstable.   

We note that the range of stability can be greatly increased by orbifolding.  The simplest group that removes all charged modes, preserves the supersymmetry, and has no fixed points (which would introduce additional light modes) is generated by $e^{2\pi i (Q_1 + 2 Q_2 - 3 Q_3)/7}$.

\subsection{Coulomb branch}

Stability here is determined by the calculation of \S4, but not assuming supersymmetric charges in evaluating the brane potential.  Using the general solution~(\ref{xsol}) in the potential terms (\ref{coul1}, \ref{coul2}) gives 
\begin{eqnarray}
{\cal E} &=&
\tilde q_K^2 +\frac1S \biggl(\tilde q_K^4 - \tilde q_K^2 \sum_I \tilde q_I^2  + |\tilde q_K| \biggr)  \nonumber\\
&=&
\frac{\tilde q_K^2 (\tilde q_I^2 + \tilde q_J^2)}{S (S + \tilde q_I^2 + \tilde q_J^2 - \tilde q_I \tilde q_J)} \bigl(\tilde q_K^2 - (\tilde q_I - \tilde q_J)^2 \bigr)
\,,\quad
I,J,K\ {\rm distinct}\,.
\end{eqnarray}
Here $S = ( \sum_I \tilde q_I^{-2})^{1/2}$.
Only the last term can change sign, so Coulomb branch is stable for all $K$ when the triangle inequality is satisfied, and unstable for some $K$ when it is violated.  This connects plausibly with the weak coupling picture of \S2, where the bosonic modes are stable if D is less than than the supersymmetric value.

\subsection {The dilaton and other instabilities}

The system appears to be have a range of stability without supersymmetry within the consistent truncation~\cite{Cvetic:1999xp,Cvetic:2000nc}, but this does not include all light modes.  Important modes not included are the $S^5$ radius and the dilaton.  We form an approximate AdS effective potential~\cite{Silverstein:2004id} for these, introducing an $S^5$ radius $r$ and dilaton $\Phi$.   The $D=3$ potential for $r$, $e^\Phi$ and $e^{U}$, in string units with order one factors omitted, is
\begin{equation}
{\cal U} \sim (e^{2U-2\Phi} r^5)^{-3} \left( - e^{2U- 2\Phi} r^3  + B^2 e^{-2U -2\Phi}  r^7 + N^2 e^{2U} r^{-5} \right) \,.
\end{equation} 
The three terms are from the $S^5$ curvature, the KK magnetic field, and the 5-form flux, and the prefactor is from the transformation to $D=3$ Einstein frame.  This potential has a minimum with $B r^2 \sim e^{2U}$ and $r^4 \sim N e^{\Phi}$, so that $r$ and $U$ are determined in terms of $\Phi$, but there is a flat direction along which $\Phi$ can be varied; in fact, in Planck units only $\Phi$ varies along this flat direction.  This was inevitable: the ingredients in our solution, and more generally those retained in the truncations, have only derivative couplings to the dilaton in Einstein frame, so that any constant value of the latter gives a solution.  

In the supersymmetic theory this flat direction should be exact: locally in the bulk there are four supersymmetries and a $U(1)$ $R$ symmetry, so no superpotential for the dilaton can arise.   In the nonsupersymmetric theories there will generically be a dilaton tadpole so that the dilaton will run.  We can still conclude that there is at least one nonsupersymmetric CFT: all ingredients are $S$-dual, so the $S$-dual point $\Phi = 0$ will be stationary.  Even if the curvature of the potential there is negative, it will be suppressed by $N$ or $\lambda$ and so be BF-stable.  More generally we could stabilize the dilaton at other values, by introducing branes and fluxes in small enough numbers to do so without affecting other modes.  Most simply, though, if we start with M theory, such as an $AdS_4 \to AdS_2 \times {\mathbb R}^2$ solution, then the dilaton will be absent from the start, while likely the other potential instabilities will have similar properties to those found here.

We note a puzzle in the supersymmetric case: the dilaton should correspond to a marginal direction, but it is not clear what this corresponds to in the CFT.  In LG models, there is not generally a line of fixed points.  It may be that the there are marginal K\"ahler potential-like terms, but these are usually expected to be irrelevant at the LG point.

Another potential instability is the screening of the magnetic field by KK monopoles.  Determining the energy for a KK monopole seems challenging, but again supersymmetry guarantees stability on a subspace.  Further we suspect that, like the perturbative magnetic instability, the stability in the supersymmetric subspace is non-marginal, and so extends at least to a neighborhood of this.  

Another instability that has arisen in some nonsupersymmetric gauge-gravity duals is the bubble of nothing~\cite{Witten:1981gj,Horowitz:2007pr}.  However, this is topologically forbidden in the supersymmetric case, even with the additional ${\mathbb Z}_k$ orbifold discussed above.  We are not modifying the topology when we break supersymmetry (at least in the limit that the magnetic plane area is infinite), so this decay should be absent.

\section{Conclusions}

The system considered here adds to the very small collection of nonsupersymmetric string states that may enjoy absolute stability, modulo the issues with the dilaton just discussed (other candidates are discussed in Ref~\cite{Narayan:2010em}).   Thus, we have a nonsupersymmetric critical system with a known UV Hamiltonian and a gravitational dual.  The $AdS_3$ region describes the physics of the LLL.  As the discussion of the entropy makes clear, the LLL inherits important interactions from the underlying gauge theory; indeed, if it were free, it could not have a gravitational dual.  We do not know these interactions precisely, but they are likely of the form $c_{ijkl}$Tr$([\lambda_i,\lambda_j][\tilde\lambda_k,\tilde\lambda_l])$, from integrating out the other components of the gauge multiplet.

One of the motivations for this work is to obtain dual examples with $AdS_2$ geometries.  Recent work from a holographic approach suggests that a coupling a Fermi sea to a CFT with $AdS_2$ dual can lead to non-Fermi liquid behavior~\cite{Lee:2008xf}.  Moreover, this appears to connect with ideas coming from condensed matter physics~\cite{Varma}.  

Such a CFT is rather exotic: it is invariant under rescaling of energy while leaving the spatial momenta fixed.  Thus its IR fluctuations are particularly strong, seemingly a necessary feature to give rise to a non-Fermi liquid.  However, it is not clear that it can be a true IR fixed point or instead is restricted to some intermediate regime (arguments suggesting the latter are given in Refs.~\cite{Jensen:2011su,Iqbal:2011in}).  To gain insight into this system, and to see whether the lessons from holography might extend to real systems, it is useful to have a holographic example in which we also know the weakly-coupled field theory.\footnote{Other approaches to $AdS_2$ include turning on a chemical potential~\cite{Chamblin:1999tk} and introducing impurity branes~\cite{Kachru:2009xf,Kachru:2010dk,Jensen:2011su}.  In both cases, avoiding the instabilities becomes a model-building challenge.}  

The solutions in this paper have a simple generalization to the $U(1)^4$ truncation of $D=11$ supergravity on $S^7$, giving $AdS_2 \times {\mathbb R}^2$ in the \mbox{IR}.  The closed form solutions are not as simple as (\ref{xsol}), but we have verified that the case $q_1 = q_2 = q_3 = +1$, $q_4 = -3$, has a supersymmetric solution while the case $q_1 = q_2 = +1$,  $q_3 = q_4 = -1$, has a nonsupersymmetric solution but no supersymmetric solution.

One of the puzzling features that must be accounted for if these duals are to be applied to real systems is the finite zero-temperature entropy of $AdS_2 \times {\mathbb R}^k$.  Ref.~\cite{D'Hoker:2009mm} suggested an intriguing field theory origin for this: the  highly degenerate LLL can be filled in many different ways, allowing for a macroscopically large entropy.  
Ref.~\cite{Almuhairi:2010rb} looked at an $AdS_5 \to AdS_2 \times {\mathbb R}^3$ example and found that the zero temperature entropy is  nonzero in the supergravity limit, but that it vanishes in the free limit.  

It seems that we can identify a weakly coupled field theory with the $AdS_2  \times {\mathbb R}^2$ by orbifolding by a ${\mathbb Z_k}$ that acts on a transverse ${\mathbb C}^2$: the dual field theory is $N$ D2-branes intersecting$k$ D6-branes.  The ratio $g^2/B$, with $g^2$ the D2 gauge coupling, is dimensionless and allows us to interpolate between weak and strong coupling.   There is a zero temperature entropy in both the strong and free limits.   The strongly coupled solution is an orbifold of the above, while in the free limit one can fill the LLL in many ways.  Simple numerology shows that these cannot match: the free field entropy always involves a factor of $\ln 2$, while the other does not.  Of course, we have seen that the interactions are crucial in the $AdS_3$ LLL, and this will be even more true in $AdS_2$ where the number of relevant interactions is much greater.  In the supersymmetric case they transmute the problem to a subtle counting of supersymmetric bound states; in the nonsupersymmetric case they would be expected to lift the macroscopic ground state degeneracy completely.  Thus, the relation between the LLL filling and the $AdS_2$ entropy is complicated.

\section*{Acknowledgments}

We are grateful to E. Silverstein for helfpul comments, to A.~Donos, J.~P.~Gauntlett and C.~Pantelidou for pointing out an error in an earlier version of this work and performing a correct stability analysis, and to F. Benini and N. Bobev for informing us of their matching of the central charges..  Part of this work was completed at the String Theory workshop of the Centro de Ciencias de Benasque; J.P. thanks many fellow participants for discussions.  This work was supported in part by NSF grants PHY05-51164 and PHY07-57035.

\begingroup\raggedright\endgroup

\end{document}